\def\simlt{\lower.5ex\hbox{$\; \buildrel < \over \sim \;$}}
\def\simgt{\lower.5ex\hbox{$\; \buildrel > \over \sim \;$}}
\RequirePackage{lineno} 
 \documentclass[preprint2]{emulateapj}

\newcommand{\myemail}{mrl@gps.caltech.edu}
\slugcomment{Submitted to the Astrophysical Journal }
\shorttitle{Assessing Chemical Disequilibrium}
\shortauthors{Line et al.}

\begin{document}
\title{A Systematic Retrieval Analysis of Secondary Eclipse Spectra III:  Diagnosing Chemical Disequilibrium in Planetary Atmospheres }
\author{Michael R. Line}
\author{Yuk L. Yung}
\affil{Division of Geological and Planetary Sciences, California Institute of Technology, Pasadena, CA 91125}
\altaffiltext{1}{Correspondence to be directed to \myemail}

\begin{abstract}
Chemical disequilibrium has recently become a relevant topic in the study of the atmospheres of of transiting extrasolar planets,  brown dwarfs, and directly imaged exoplanets.   We present a new way of assessing whether or not a Jovian-like atmosphere is in chemical disequilibrium from observations of detectable or inferred gases such as H$_2$O, CH$_4$, CO, and H$_2$.  Our hypothesis, based on previous kinetic modeling studies, is that cooler atmospheres will show stronger signs of disequilibrium than hotter atmospheres.  We verify this with chemistry-transport models and show that planets with temperatures less than $\sim$1200 K are likely to show the strongest signs of disequilibrium due to the vertical quenching of CO, and that our new approach is able to capture this process.  We also find that in certain instances a planetary composition may appear in equilibrium when it actually is not due to the degeneracy in the shape of the vertical mixing ratio profiles.  We determine the state of disequilibrium in eight exoplanets using the results from secondary eclipse temperature and abundance retrievals.  We find that all of the planets in our sample are consistent with thermochemical equilibrium to within 3-sigma.  Future observations are needed to further constrain the abundances in order to definitively identify disequilibrium in exoplanet atmospheres.
\end{abstract}

\section{Introduction}
Disequilibrium mechanisms play an important role in determining the atmospheric composition of planets and cool substellar objects.   For instance, Prinn \& Barshay (1977) showed that eddy mixing could explain the anomalously high abundance of CO in Jupiter's observable atmosphere by dredging up CO-rich gas from the hotter, deeper atmosphere.    Photochemistry significantly alters the stratospheric compositions of all planetary atmospheres in our solar system by enhancing or depleting various species.  Specifically, methane photolysis is responsible for the production of heavy hydrocarbons in the gas giant atmospheres (Yung \& DeMore 1999).  Zonal winds can transport photolytically produced species from the dayside to the nightside.  It is not unreasonable to assume that these mechanisms play an equally important role in exoplanet atmospheres.  In fact there have been some observational suggestions of disequilibrium in exoplanet atmospheres (Stevenson et al. 2010), although this conclusion has recently been called into question (Line et al. 2011; Moses et al. 2013). 

Motivated by recent exoplanet atmosphere observations, a variety of 1D photochemical-transport models have been used to explore the compositions of hot Jupiters (Liang et al. 2003, 2004; Zahnle 2009a,2009b; Line et al. 2010; Moses et al. 2011; Visscher \& Moses 2011; Kopparapu et al. 2012; Venot et al. 2012), hot Neptunes (Line et al. 2011; Moses et al. 2013), and super Earths (Miller-Ricci Kempton et al. 2012; Hu et al. 2012); some 3-D chemistry-transport models have been used in an analogous way to study hot Jupiters (e.g. Cooper \& Showman 2005; Ag{\'u}ndez et al. 2012). The basic conclusion from most of these studies is that in cool atmospheres (T $\le$ 1200 K), as chemical reaction timescales increase, disequilibrium mechanisms become increasingly more important while in hot atmospheres thermochemical equilibrium prevails. 

From these investigations, we have also learned that Jovian-like planetary atmospheres can be vertically divided into three basic chemical regimes.   In the deep atmosphere where temperatures and pressures are high, chemical reaction timescales are short, allowing the composition to achieve thermochemical equilibrium.  At lower pressures and temperatures higher up in the atmosphere, chemical reaction timescales slow until the point at which they are equal to the vertical transport timescale, thus quenching the abundances.  Vertical mixing tends to smooth out the vertical mixing ratio profiles.  At even higher altitudes in the atmosphere, ultraviolet photons can break apart stable molecules and alter the upper atmospheric composition. 

Chemical disequilibrium models demonstrate that vertical transport tends to have the greatest observational consequence for exoplanet atmospheres because the infrared photosphere of most exoplanets (with current instruments) tends to fall within the region of the atmosphere dominated by vertical mixing but not yet strongly affected by photochemistry.    Therefore we would expect to observe a trend showing that the compositions of cooler atmospheres deviate strongly from thermochemical equilibrium, and hot atmospheres to remain in thermochemical equilibrium.  We seek to answer the simple ``yes or no" question of ``Is this atmosphere in chemical disequilibrium?"  In \S \ref{sec:theory} we describe a simple way of looking at planetary compositions in order to diagnose chemical disequilibrium and to seek a trend in the strength of disequilibrium with temperature.  In \S \ref{sec:model} we show how our method compares with 1-D chemical transport models. \S \ref{sec:results} presents our diagnosis of disequilibrium for eight exoplanets using the temperature and abundance retrieval results from Paper II (Line et al. 2013). 


\section{Theory}\label{sec:theory}
In order to address our hypothesis, we seek a quantity that relates a planet's composition to its temperature.  Given some measurement of the abundances of various gases, one could try to determine if any one of those gases are in or out of equilibrium.  However, looking at individual gases is difficult because their abundances depend on the planetary elemental abundances as well as temperature.       Therefore, we seek a relationship that relates composition to temperature in a way that is independent of metallicity and the C/O ratio.   That quantity can be derived as follows.
In thermochemical equilibrium the net reaction

\begin{equation}\label{eq:equation5_1}
CH_4+H_2O=CO+3H_2
\end{equation}
relates the abundances of CH$_4$, CO, and H$_2$O, and H$_2$.  We choose these species because, generally, in extra-solar planet and substellar object atmospheres, these species are the most abundant and readily detectable or inferred by infrared telescope facilities (Tinetti et al. 2007; 2010; Grillmair et al. 2007; 2008;  Swain et al. 2009a; 2009b).   From the law of mass action we have the relation
\begin{equation}\label{eq:equation5_2}
	\alpha(f_{i},P)=\frac{f_{CH_4}f_{H_2O}}{f_{CO}f_{H_2}^3P^2}=K_{eq}(T)
\end{equation}
where we have defined $\alpha(f_{i},P)$ to be the combination of the gas mixing ratios and pressure,  $f_{i}$ is the mixing ratio of species $i$,  $P$ is the pressure at some specified level in the atmosphere (in bars), and  $K_{eq}(T)$ is the equilibrium constant at temperature $T$ (Yung \& DeMore 1999).   The equilibrium constant only depends on temperature and the thermodynamic properties of the molecules and generally has the form
\begin{equation}\label{eq:equation5_3}
	K_{eq}(T)=e^{-\Delta G/RT}= e^{-(\Delta H/RT-\Delta S/R)}
\end{equation}
 where $\Delta G$, $\Delta H$ and $\Delta S$ are the change in Gibbs free energy, enthalpy and entropy, respectively, of the molecules involved.    These quantities can be found in any thermodynamic table (e.g., the NASA ThermoBuild website\footnote{ http://www.grc.nasa.gov/WWW/CEAWeb/ceaThermoBuild.htm}).  From Equation \ref{eq:equation5_2}  we see that if we can measure or infer the abundances of CH$_4$, CO, H$_2$O, and H$_2$ at a known pressure level we can relate them to a quantity that solely depends on temperature and the thermodynamic properties of each molecule. 
 
\begin{figure}[h]
\begin{center}
\includegraphics[width=0.5\textwidth, angle=0]{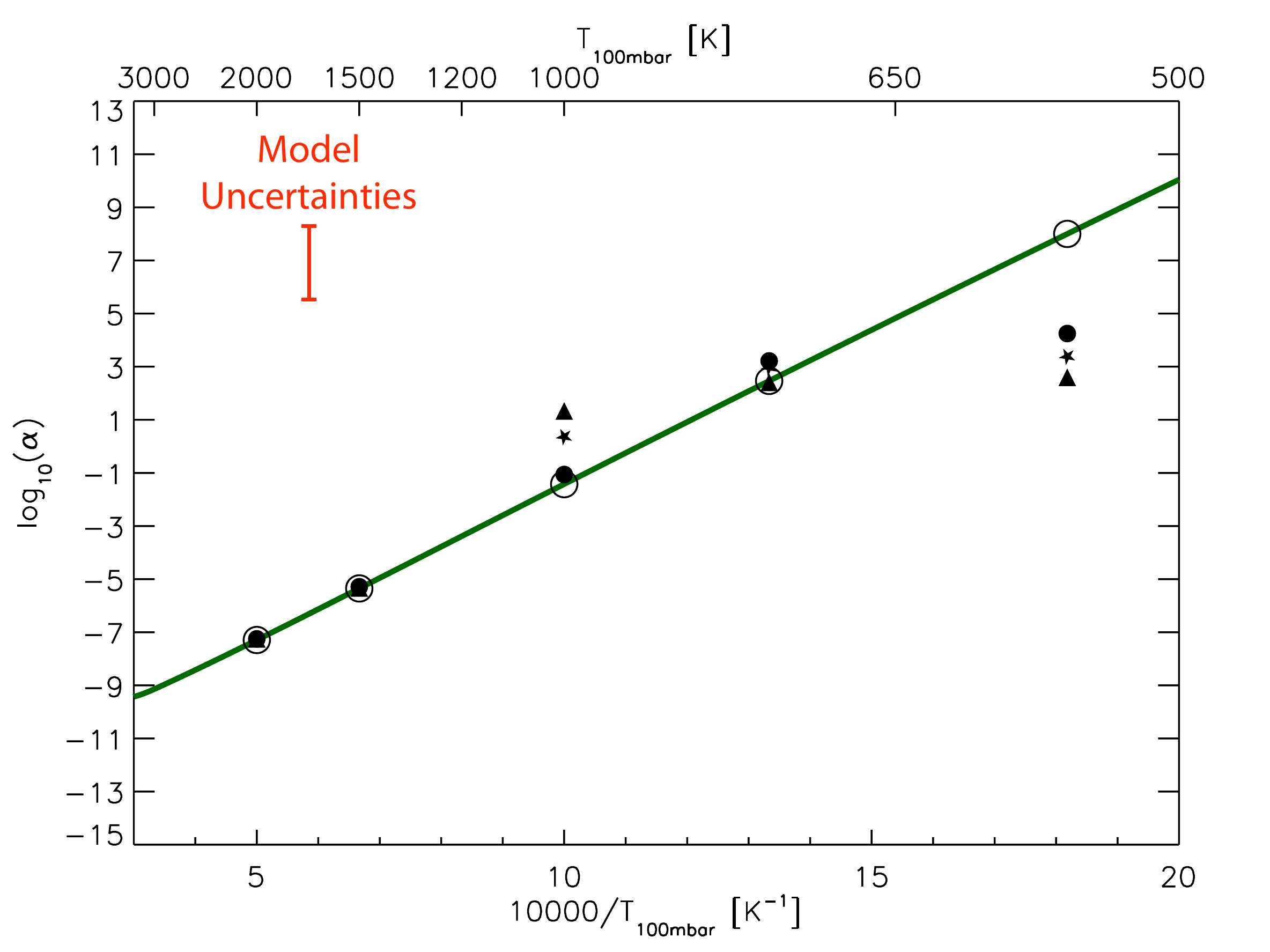}
\end{center}
  \caption[Equilibrium Constant Plot with Model]{ \label{fig:figure2} Log of the equilibrium constant as a function of 1/ temperature ($K_{eq}(T)$, green) compared with vertical transport models.  Each point represents an evaluation of both sides of Equation \ref{eq:equation5_2} at the 100 mbar level from the models in Figure \ref{fig:figure3}.  The open circles represent these values for atmospheres in thermochemical equilibrium.  The solid circles represent the atmospheres under an eddy diffusion coefficient of 10$^7$ cm${^2}$s$^{-1}$, the star, an eddy diffusion coefficient of 10$^9$ cm${^2}$s$^{-1}$, and the triangle with an eddy diffusion coefficient of 10$^{11}$ cm${^2}$s$^{-1}$.  At hot temperatures all four points fall on top of each other, suggesting that the atmosphere is in thermochemical equilibrium at the 100 mbar pressure level.  The model uncertainty introduced (red) due to the uncertainty in the H$_2$ mole fraction and sensed pressure level is also shown.    }
\end{figure} 
 
Figure \ref{fig:figure2} shows the equilibrium constant (green curve) in Equation \ref{eq:equation5_2} as a function of temperature.  In equilibrium, $\alpha$ at a specified pressure, is equal to the equilibrium constant.  We choose to evaluate $\alpha$ at the 100 mbar pressure level.  This pressure level is where most secondary eclipse thermal emission weighting functions tend to peak (e.g.,  Line et. al. 2013), and hence temperature and abundance determinations sample this region.  If we determine the abundances of the aforementioned gases and find that $\alpha$ has the same value as the equilibrium constant, K$_{eq}$, evaluated at the 100 mbar temperature, then we might infer that those four gases are in thermochemical equilibrium.  \footnote{There is a caveat here.  There are some situations in which the veritcal mixing of species can result in a combination of abundances that can mimic equilibrium at a given pressure level.  This is discussed in more detail in the next section} This is equivalent to the value of $\alpha$ falling on the line in Figure \ref{fig:figure2}.  If however, $\alpha$ is not equal to the equilibrium constant at that temperature, then we can infer that the four gases are not in thermochemical equilibrium and that there must be some process driving those species away from equilibrium.   For instance, as we will show in the next section,  for cool atmospheres if vertical mixing is operating, CO will be dredged up from deeper, more CO rich regions thus causing $\alpha$ to be less than the equilibrium constant value.   In this investigation we again simply choose to focus on CH$_4$, CO, H$_2$O, and H$_2$, but in principle, any set of gases can be related to an equilibrium constant.  
  
 We note that in most instances, H$_2$ is not readily spectroscopically constrained.  It can, however, be determined through mass balance arguments by assuming the sum of the mixing ratios of all species must be unity and that H$_2$ or some known or assumed combination of H$_2$ and other spectroscopically inactive gases are the only filler gases.  Additionally, through interior modeling of the mass-radius relationship of a planet/substellar object we may also be able to put reasonable constraints on the bulk H$_2$ abundances (Fortney et al. 2007; Baraffe et al. 2008;  Rogers et al. 2011; Mordasini et al. 2012).     We will explore the impact of the uncertainties in the H$_2$ in the discussion section.  In any case, the key is to choose gases that can be constrained to within reason with current observational capabilities.

\begin{figure*}
\begin{center}
\includegraphics[height=4.5in,width=!, angle=0]{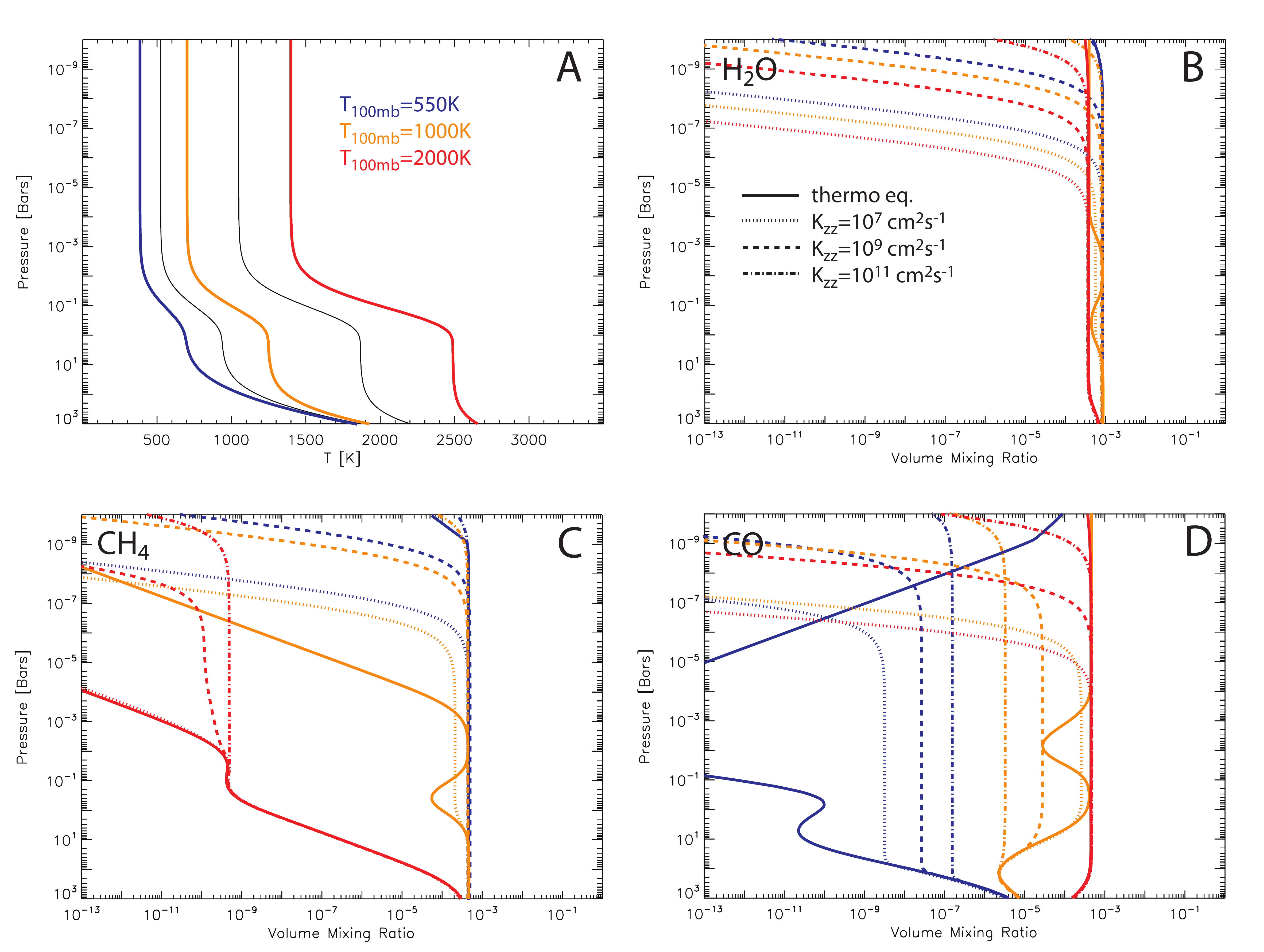}
\end{center}
     \caption[Mixing Ratio Vertical Profiles]{ \label{fig:figure3} Model atmospheres subject to a variety of vertical mixing strengths (K$_{zz}$).  The upper left panel shows 5 different temperature profiles with 100 mbar temperatures of 550, 750, 1000, 1200, 1500, and 2000 K.  In the subsequent panels, for each colored temperature profile, the corresponding vertical mixing ratio profiles are shown for H$_2$O (top right), CH$_4$ (bottom left) and CO (bottom right) under different vertical mixing strengths.  The thermochemical equilibrium derived mixing ratio profiles are solid, while the dotted mixing ratio profiles are derived with an eddy diffusion coefficient of 10$^7$ cm${^2}$s$^{-1}$, the dashed with an eddy diffusion coefficient of 10$^9$ cm${^2}$s$^{-1}$, and the dot-dashed with an eddy diffusion coefficient of 10$^{11}$ cm${^2}$s$^{-1}$ }
\end{figure*} 
 
\section{Approach Validation with Vertical Transport Models}\label{sec:model}
We use a 1-D chemical kinetics model (Allen et al. 1981) modified for exoplanets (Line et al. 2010;  2011) to explore the consequences of vertical mixing on the relationship established in Equation \ref{eq:equation5_2} as a function of temperature.  We do not focus on photochemistry in this investigation as it has been shown that the observational consequence on these species is minimal (Moses et al. 2011) because of the relatively low abundances of optically active photochemical by-products in the deep photospheric region we are interested in.  The model can kinetically achieve thermochemical equilibrium in the deep atmosphere and seamlessly transition across the different chemical regimes  (see Visscher \& Moses 2011 for an in-depth discussion on this topic).  We use the reaction database of Line et al. 2011 which includes the important reactions in dictating the quench levels of CH$_4$ and CO (e.g., Visscher \& Moses 2011; Moses et al. 2011).  We generate a series of exoplanet temperature-pressure profiles under different levels of irradiation to produce a range of effective temperatures (given as 100 mbar temperature) using the Guillot (2010) analytic temperature relation (Figure \ref{fig:figure3}a).

For each of the temperature profiles we first compute the thermochemical equilibrium composition under the assumption of solar \footnote{While Equation \ref{eq:equation5_2} will be valid in equilibrium regardless of the elemental abundances , the elemental abundances may play a role in the disequilibrium processes.  That is, different metallicities or C/O ratios could potentially change the location of the quench levels and quench level abundances} elemental abundances (solid curves in Figures \ref{fig:figure3}b,c, and d).   For this we use the NASA Chemical Equilibrium with Applications model (Gordon \& McBride 1996).  Next we compute the disequilibrium compositions under the assumptions of different vertical mixing strengths.   The vertical mixing strength is parameterized via eddy diffusion (K$_{zz}$).  The strength of eddy diffusion in exoplanet atmospheres is not well known but can be estimated with a mixing length theory (e.g., Line et al. 2010, Moses et al. 2011, Line et al. 2011) using vertical wind profiles derived from General Circulation Models (Showman et al. 2009).  This generally gives an order-of-magnitude estimate for the eddy mixing strengths.    For reference, the eddy diffusion strength in Jupiter is thought to span 10$^7$-10$^9$ cm$^2$s$^{-1}$ (Prinn \& Barshay 1977) in order to explain the anomalous stratospheric CO abundance.  A similar investigation by Griffith \& Yelle (1999) estimate brown dwarf eddy diffusion strengths to be on the order of 10$^6$ cm$^2$s$^{-1}$.  For simplicity, we assume constant-with-altitude eddy mixing profiles that span a reasonable range of plausible values from 10$^7$-10$^{11}$ cm$^2$s$^{-1}$ (dashed and dotted curves in Figures \ref{fig:figure3}b,c, and d).   

The vertical mixing ratio profiles all share some general features under vertical mixing.  Deep in the atmosphere, where temperatures are high and reaction timescales are short, the profiles converge towards thermochemical equilibrium.  As temperatures cool, vertical mixing dominates, smoothing out the profiles to a constant-with-altitude vertical structure.  At the highest region of the atmosphere molecular diffusion dominates, causing a rapid fall-off in the mixing ratios.

For water (Figure \ref{fig:figure3}b) we find that there is little effect on the vertical composition due to vertical mixing or temperature.  Therefore water is not a good tracer for chemical disequilibrium, at least in the hot-Jovian-like atmosphere regime (though it may be in high C/O regimes where its abundance is less).   Methane (Figure \ref{fig:figure3}c) however, is more strongly affected by vertical mixing.  At cool temperatures (blue) the vertical profile of methane is unaffected by disequilibrium due to its overwhelmingly large abundance, but is more strongly affected at higher temperatures because it becomes a trace species.  At low eddy diffusion strengths, however, the chemical equilibrium timescales overcome the eddy mixing strength to achieve thermochemical equilibrium.  If we were to go to even higher temperatures, say, T$_{100mbar}$=2500 K,  methane would maintain equilibrium throughout.  At cool temperatures, CO is the molecule most affected by vertical mixing.  At cool temperatures (blue) it is clear that vertical mixing can result in orders-of-magnitude changes in the CO abundances in the infrared photosphere.  In the deep atmosphere, CO achieves thermochemical equilibrium, but readily moves towards disequilibrium near 100 bars.  An increase in the eddy diffusion strength at these cool temperatures results in an increase in the disequilibrium CO abundance in the infrared photosphere due to the shape of the thermochemical profile in the deep atmosphere.     At warm temperatures (orange) the reverse occurs. The shape of the thermochemical profile of CO changes, resulting in a decrease in the CO abundance with increasing mixing strength.  And finally, at hot temperatures (red), there is virtually no effect of from vertical mixing in the infrared photosphere region.   

If for each of the models shown in Figure \ref{fig:figure3} we evaluate $\alpha$ at 100 mbars as a function of temperature at 100 mbars we can place a point on Figure \ref{fig:figure2}.   We find that indeed, for the hottest planets, (T$_{100mb}$ $>  \sim$1200 K ), that CH$_4$, CO, H$_2$O, and H$_2$ are in thermochemical equilibrium (they fall on the thermochemical equilibrium line) even under a wide range of vertical mixing strengths.  At the coolest temperatures, models subject to vertical mixing begin to fall below the thermochemical equilibrium line due to the vertical transport of CO rich gas.   

Since the vertical structure of H$_2$O and H$_2$ are generally independent of temperature we can better understand the ratio in Equation \ref{eq:equation5_2}, and where points will fall in Figure \ref{fig:figure2} by looking at the ratio of CH$_4$/CO which is given as
\begin{eqnarray}\label{eq:equation5_4}
	\frac{f_{CH_4}}{f_{CO}} \propto P^2K_{eq}(T(P))= P^2e^{-(\Delta H/RT(P)-\Delta S/R)}\nonumber\\ 
	\approx P^2e^{27086/T(P)-30} 
\end{eqnarray}
From Equation \ref{eq:equation5_4} we see that the detailed vertical structure of these two species will strongly depend on the functional form of $T(P)$.  A high CH$_4$ abundance will be favored at high pressures and low temperatures.  Low temperatures at high pressures are favored in overall cooler temperature structures.   Conversely, CO will be favored at low pressures and high temperatures which occur in overall hotter temperature structures.   In isothermal regions of the atmosphere, decreasing pressures will favor an increasing CO abundance.  

Figure \ref{fig:figure2} shows some ambiguity at 100 mbar temperatures between $\sim$700 and $\sim$1100 K.  Because of the structure of the temperature profile and how that plays into Equation \ref{eq:equation5_3}  we find that near T$_{100mbar}$$\sim$750 K the CO thermochemical profile (not shown) is such that the mixing ratios near the quench level happen to be nearly identical to the mixing ratios at 100 mbars.   So, even though there is disequilibrium in this region of the atmosphere, there is also a degeneracy in the thermochemical mixing ratio profile.   This explains why, in Figure \ref{fig:figure2}, most of the points at T$_{100mbar}$=750 K seem to fall on, or very nearly on, the thermochemical equilibrium line.   Rather than claiming equilibrium when $\alpha$ falls on the line, we would be safer to say that if $\alpha$ falls off of the line it is in disequilibrium.  The temperature structure in the T$_{100mbar}$=1000 K profile interacts with Equation \ref{eq:equation5_3} in such a way that near the CO quench level the slope of the CO mixing ratio profile (Figure \ref{fig:figure3},d) is positive.  This is because the temperature at the CO quench levels is isothermal resulting in a $1/P^2$ dependence in the CO mixing ratio.  This behavior does not occur in the less irradiated cooler profiles because the quench levels occur deeper in the region of the atmosphere where the temperature is always decreasing with altitude (faster than the $1/P^2$ dependence) resulting in a decreasing CO  mixing ratio with altitude.   

 We also point to two caveats.  In Jovian-type exoplanets planets we cannot spectroscopically determine the H$_2$ mixing ratio at this time (e.g., Lee et al. 2012).  Additionally, we may not necessarily know the pressure level at which we are sensing.  Through a Monte Carlo analysis we explore the effect that the uncertainties in the bulk H$_2$ abundance and the assumed pressure level have on our ability to determine disequilibrium.  Generally, if we are exploring hot-Jovian type objects, the H$_2$ abundance will remain near $\sim$80\%.  However,  we choose a liberal range of values from 10\% to 100\% over which we generate a uniform distribution.  We would expect interior modeling (Fortney et al. 2007; Baraffe et al. 2008;  Rogers et al. 2011; Mordasini et al. 2012) of extra-solar planets and substellar object to at least give us an estimate within this range.   We have also assumed that most observations originate from the 100 mbar pressure level.  This may not always be the case.  We do believe, however, that most emission from such objects will emanate between 1 bar and 10 mbars.  As such, we generate a logarithmically uniform distribution of pressures between these values.   With fixed values of CO, CH$_4$, H$_2$O, and many combinations of pressures and hydrogen mole fractions we recompute $\alpha$.  The spread in values, represented by the standard deviation of the resultant distribution,  is shown in Figure \ref{fig:figure2}.  We find the overall uncertainty due to the location of the pressure level and H$_2$ mole fraction is much smaller than current observational capabilities, and hence is not a limitation.  As observational capabilities improve, this ``model" uncertainty becomes more significant relative to the observational uncertainties.  However, at the point at which our spectroscopic capabilities are that refined, we should be able to more accurately pin down the H$_2$ mole fraction, especially with proposed missions like CHEOPS (Broeg et al. 2013) and TESS (Ricker et al. 2009).  Also, with higher spectroscopic resolution we can be more certain where various absorption bands are and hence more accurately locate at which pressure levels the emission from a certain molecule is coming from.  With current observational technology, dominated by broadband measurements, it is difficult to resolve any vertical abundance information (e.g., Lee et al. 2012).  


\section{Observational Consensus of Disequilibrium in Exoplanet Atmospheres}\label{sec:results}
\begin{figure*}[h]
\begin{center}
\includegraphics[width=0.85\textwidth, angle=0]{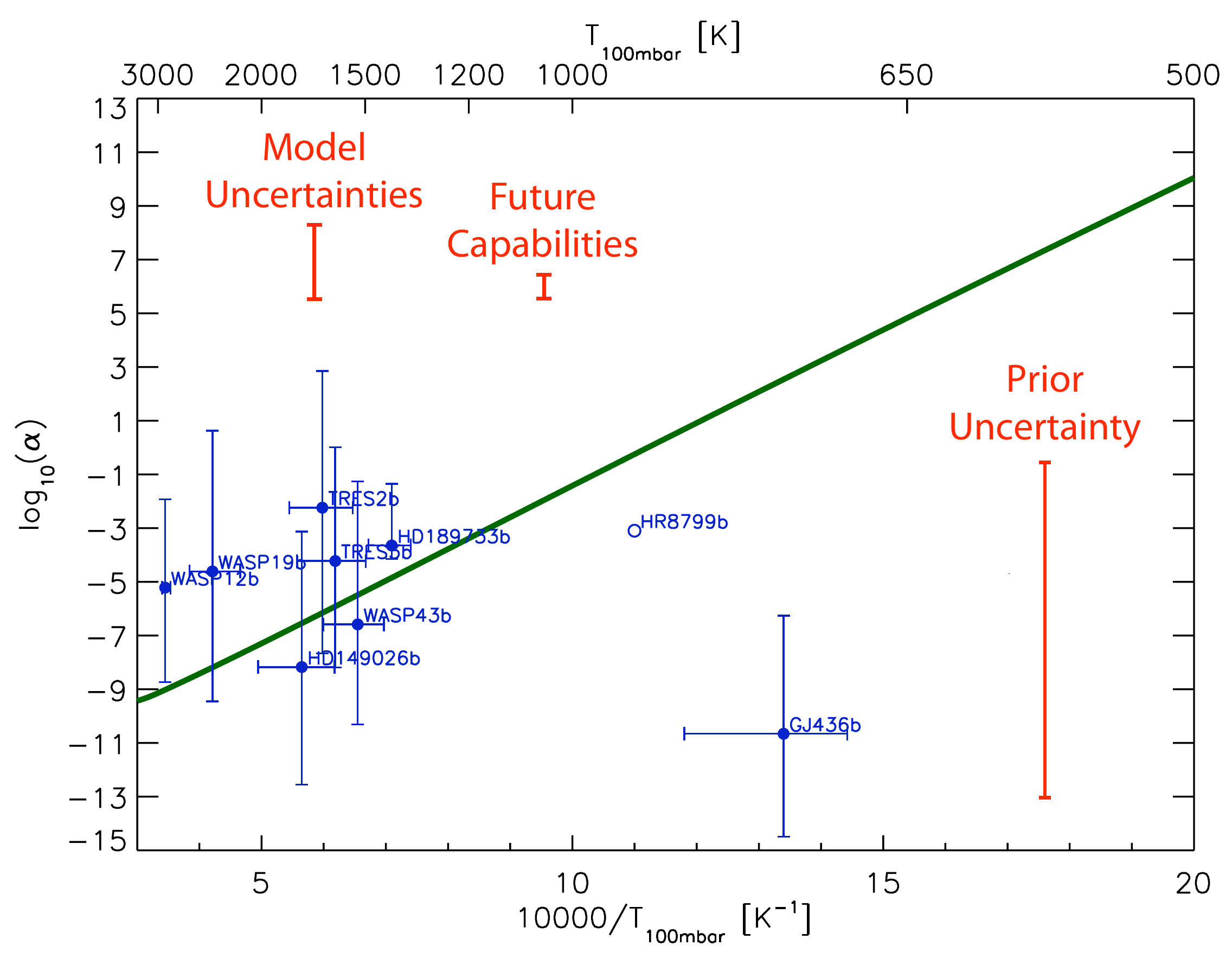}
\end{center}
     \caption[Mixing Ratio Vertical Profiles]{ \label{fig:figure7} Assessment of disequilibrium from secondary eclipse data. The green curve is the log of the equilibrium constant as a function of 1/temperature ($K_{eq}(T)$).  The blue points with error bars are an evaluation of $\alpha$ based on the retrieval analysis.  Values of alpha that are consistent with the green line are consistent with equilibrium.  Also shown are the model uncertainties, prior uncertainties, and future observational capabilities (red).   The open blue circle are the recent best fit retrieval results for the direct imaged planet HR8799b from Lee et al. 2013.    }
\end{figure*}

Using the temperature and abundance retrieval results from Paper II (Line et al. 2013) we assess disequilibrium in the atmospheres of HD189733b, HD149026b, GJ436b, WASP-12b, WASP-19b, WASP-43b, TrES-2b, and TrES-3b. By propagating the uncertainties from the Markov chain Monte Carlo analysis in Part II,   we can compute an $\alpha$ probability distribution at the 100 mbar pressure level.  From this $\alpha$ probability distribution we can compute the median and 68$\%$ error bar.  With the temperature profile distribution we can compute the 68$\%$  uncertainty level of the temperature at 100 mbars.   These results are shown in Figure \ref{fig:figure7}.  When interpreting the error bars we must keep in mind that we imposed upper and lower limits on the DEMC gas prior.   In order to facilitate the interpretation of the derived uncertainties in $\alpha$, we show the 68$\%$ error bar resulting from imposed upper and lower limits described above.  We find that the uncertainties in $\alpha$ on each planet are less than this prior uncertainty.  This means we obtained at least some information from the observations.  Also, we show again the model uncertainties due to our lack of knowledge of the exact pressure level of the observations and H$_2$ mixing ratio.  One can fold this uncertainty in by adding in quadrature the model uncertainty to the data uncertainties.  We have also included the best fit retrieval results of direct imaged planet HR8799b from Lee et al. 2013.  

From this analysis we find that all of the planets are consistent with thermochemical equilibrium within $\sim$3-sigma.  GJ436b is the most likely candidate to be in disequilibrium.  The value of $\alpha$ falls almost exactly 3-sigma blow the equilibrium line.  However, if we fold in the model uncertainty the disequilibrium detection significance is less.  The large deviation from equilibrium is due to the high abundance of CO required to fit the lack of flux in the 4.5 $\mu$m bandpass and the low CH$_4$ mixing ratio required to fit the 3.6 $\mu$m flux.  This high CO and low CH$_4$ abundance as described in Moses et al. 2013 is due to the possible high (300x) metallicity of GJ436b.  Though increasing the metallicity alone to this level does not fully reproduce the 3.6 to 4.5 $\mu$m flux contrast.  It is also possible that we are not accounting for unknown absorbers since we currently only include four in our retrieval model.  The value of $\alpha$ for HR8799b also falls off of the equilibrium curve but we have no estimate of the uncertainties so we cannot assess the significance of this detection.  

The large uncertainties on the $\alpha$ stress the need for significantly better data in order to truly determine chemical equilibrium or disequilibrium in an exoplanetary atmosphere. In Figure \ref{fig:figure7} we show future observational capabilities based on the dayside emission spectra synthetic retrieval results of a FINESSE-like instrument (Part I, Line et al. 2013a) to demonstrate the future potential to identify disequilibrium.   Such a FINESSE-like telescope (Swain et al. 2012), would be able to identify disequilibrium to $\sim$2 or $\sim$3 sigma especially in the cooler planets where the effects should be the strongest.

\section{Discussion \& Conclusions}
We have developed a simple way of determining if a planetary or substellar object atmosphere is subject to disequilibrium mechanisms (Figure \ref{fig:figure2}).  We have chosen CO, CH$_4$, H$_2$O, and H$_2$ as our tracer species as they are the most abundant radiatively active or readily inferred gases in a variety of observationally accessible planetary and substellar object atmosphere environments.    If we determine the abundances of CO, CH$_4$, H$_2$O, and H$_2$ we can assess whether these species are subject to some form of disequilibrium.  If we observe that the ratio of these four species in Equation \ref{eq:equation5_2} is not equal to the equilibrium constant at the observed temperature, and hence does not fall on the equilibrium line in Figure \ref{fig:figure2},  then we can safely conclude that there is a process driving them out of equilibrium.  If however, the observed ratio is consistent with equilibrium, we may conclude that either the atmosphere is indeed in equilibrium, or that the disequilibrium process interacts with the temperature-pressure profile in such a way as to make it appear as if the planetary atmosphere were in equilibrium. 

We also note that in this investigation we used vertical mixing as our example disequilibrium mechanism.   There are other process such as photochemistry, biology, cloud formation, etc.  For instance, in some extreme cases, photochemistry may drive the carbon out of CH$_4$ and into HCN.  If the photochemistry is vigorous enough to deplete methane over the infrared photosphere (unlikely), we might expect to find the planet to fall below the thermochemical equilibrium line.  If we were to observe planets as cool as the Jovian's in our solar system, we would find that they fall far below the thermochemical equilibrium line because of the depletion of water due to condensation.  So while we can determine if there is some disequilibrium process occurring, it may be difficult to disentangle what that process is, but simply identifying disequilibrium would be exciting enough to warrant future observations of that planet.

Using the temperature and abundance retrieval results from Paper II,  we found that all of the planets within our sample were consistent with equilibrium to within 3-sigma.  This is due to the large uncertainties on the gas abundances.  Future observations are needed in order to definitively identify disequilibrium in exoplanet atmospheres.  Furthermore, we need more high quality observations of cool (500-1200 K) planets since these planets are the most likely to show signs of chemical disequilibrium.

\section{Acknowledgements}		
We thank Dave Stevenson, Pin Chen, Gautam Vasisht, and Channon Visscher for useful discussions.  We also thank Julie Moses and Jonathan Fortney as well as the Yuk Yung group for meticulously reading the manuscript.  This research was supported in part by an NAI Virtual Planetary Laboratory grant from the University of Washington to the Jet Propulsion Laboratory and California Institute of Technology. YLY was supported in part by NASA NNX09AB72G grant to the California Institute of Technology.

%


\end{document}